\documentclass[usenatbib]{mn2e}
\usepackage{natbib}
\usepackage{times}
\usepackage{graphicx}

\newcommand{\beq}{\begin{equation}}
\newcommand{\eeq}{\end{equation}}
\newcommand{\beqnn}{\begin{displaymath}}	
\newcommand{\eeqnn}{\end{displaymath}}		
\newcommand{\beqa}{\begin{eqnarray}}
\newcommand{\eeqa}{\end{eqnarray}}
\newcommand{\beqann}{\begin{eqnarray*}}
\newcommand{\eeqann}{\end{eqnarray*}}

\newcommand{\ben}{\begin{enumerate}}
\newcommand{\een}{\end{enumerate}}
\newcommand{\bit}{\begin{itemize}}
\newcommand{\eit}{\end{itemize}}
\newcommand{\bc}{\begin{center}}
\newcommand{\ec}{\end{center}}

\newcommand{\eqref}[1]{Equation~\ref{#1}}

\def\la{\mathrel{\hbox{\rlap{\hbox{\lower4pt\hbox{$\sim$}}}\hbox{$<$}}}}
\def\ga{\mathrel{\hbox{\rlap{\hbox{\lower4pt\hbox{$\sim$}}}\hbox{$>$}}}}

\def\fdg{\hbox{$.\!\!^\circ$}}

\def\laeq{\lower.5ex\hbox{{$\:\scriptstyle\buildrel < \over \sim\:$}}}
\def\gaeq{\lower.5ex\hbox{{$\:\scriptstyle\buildrel > \over \sim\:$}}}

\makeatletter 
 \def\sub#1{\relax\ifmmode _{\fam\z@ #1}\else
         $_{\fam\z@ #1}$\fi}
 \def\super#1{\relax\ifmmode ^{\fam\z@ #1}\else
         $^{\fam\z@ #1}$\fi}
\makeatother

\newcommand{\mycaption}[3]
{\if*#2 \caption{#3\label{#1}}
 \else  \caption[#2]{#3\label{#1}}
 \fi}

\newcommand{\comment}[1]{\relax}

\long\def\COMMENT#1\ENDCOMMENT{}
\def\ENDCOMMENT{}

\newcommand{\bhyi}{\mbox{$\beta$~Hyi}}

\def\oversim#1#2{\lower0.5pt\vbox{\baselineskip0pt \lineskip-0.5pt
     \ialign{$\mathsurround0pt #1\hfil##\hfil$\crcr#2\crcr\sim\crcr}}}

\title[Radius and mass of \bhyi]
{The radius and mass of the subgiant star \bhyi{} from
  interferometry and asteroseismology}
\author[J. R. North et al.]
       {
J. R. North,$^1$
J.    Davis,$^1$
T. R. Bedding,$^1$
M. J. Ireland,$^{1,2}$
A. P. Jacob,$^1$
J.    O'Byrne,$^1$\newauthor
S. M. Owens,$^1$
J. G. Robertson,$^1$ 
W. J. Tango$^1$ and
P. G. Tuthill$^1$
\\
	$^1$School of Physics, University of Sydney 2006, Australia\\
	$^2$Planetary Science, MS 150-21, Caltech, 1200 E. California Blvd, Pasadena, CA 91125, USA
}

\begin{document}

\maketitle

\begin{abstract} 
We have used the Sydney University Stellar Interferometer (SUSI) to measure
the angular diameter of $\beta$~Hydri.  This star is a nearby G2 subgiant
whose mean density was recently measured with high precision using
asteroseismology.  We determine the radius and effective temperature
of the star to be $1.814\pm0.017\,{\rm R}_\odot$ (0.9\%) and 
$5872\pm44$\,K (0.7\%) respectively. By combining this value with the mean
density, as estimated from asteroseismology, we make a direct estimate of
the stellar mass.  We find a value of $1.07 \pm 0.03\,{\rm M}_\odot$
(2.8\%), which agrees with published estimates based on fitting in the H-R
diagram, but has much higher precision. These results place valuable
constraints on theoretical models of $\beta$~Hyi and its oscillation
frequencies.
\end{abstract}


\section{Introduction}

The combination of interferometry and asteroseismology is a powerful way of
constraining the parameters of stars
\citep*{KTS2003,PTG2003,KTM2004,TKP2005,vBCB2007,CMM2007}.  Interferometry provides
an angular diameter which, combined with the parallax, gives a direct
measurement of the stellar radius.  Asteroseismology, on the other hand, uses the
oscillation frequencies of the star to infer details about its internal
structure (e.g., \citealt{B+G94}).  In particular, $\Delta\nu$, the
so-called large frequency separation between consecutive radial overtones,
gives a good estimate of the mean stellar density.  The combination of the
radius and $\Delta\nu$ can therefore provide a direct measurement of the
mass of a star.

The star \bhyi{} (HR 98, HD 2151, HIP 2021) is a southern evolved subgiant
(spectral type G2~IV) with a mass slightly greater than the Sun and an age of
about 6.5--7.0\,Gy \citep{DLvdB98,F+M2003}. This star is an excellent target 
for asteroseismology, being the closest G-type subgiant. Its distance, 
thanks to {\em Hipparcos\/}, is known to 0.36\% accuracy. It is bright ($V=2.8$) and 
has a low $v\sin i$, which allows extremely good Doppler precision.  
Solar-like oscillations in \bhyi{} were recently measured by \citet{BKA2007}, 
who used the large frequency separation to infer the mean stellar density to 
an accuracy of 0.6\%.

In this paper we report the first measurement of the angular diameter of
\bhyi.  We use the {\em Hipparcos\/} parallax to infer the radius, which we
combine with the mean density, determined from asteroseismology, to
estimate the mass of the star. The effective temperature and surface gravity
are also constrained using our radius and complementary data from the
literature.

%
%

\section{Observations and Data Reduction}
\label{obs}

The Sydney University Stellar Interferometer (SUSI, \citealt{DTB99}) was 
used to measure the squared visibility (i.e.\ normalised squared modulus of 
the complex visibility) or $V^2$ on a total of 7 nights. The 
red-table beam-combination system was employed with a filter of centre 
wavelength and full-width half-maximum 700\,nm and 80\,nm respectively. 
This system is to be described in greater detail by Davis et al.\ 
(in preparation) and an outline is given by \citet{TDI2004}.

The interference produced by a pupil-plane beam-combiner was 
modulated by repeatedly scanning the optical delay about the white
light fringe position. Two avalanche photo-diodes detect the two outputs
of the combining beamsplitter. An observation consisted of a set of 1000 
scans, each traversing 140\,$\mu$m in optical delay, sampled in 1024 
steps of 0.2\,ms duration.

The post-processing software mitigates the effect of scintillation in each
observation by differencing the two recorded signals. The squared visibility 
was estimated after bias subtraction and a nonlinear function was applied to 
partially correct for residual seeing effects \citep{Ire2006}. 

Target observations were interleaved with calibration measurements of nearby 
stars chosen to provide a known or essentially unresolved signal. The angular 
diameters (and associated error) of these calibrators stars were estimated 
from an intrinsic colour interpolation (and spread in data) of measurements 
made with the Narrabri Stellar Intensity Interferometer \citep{HBDA74},
including corrections for the effects of limb-darkening.
Using the adopted stellar parameters of the calibrator stars given in 
Table~\ref{cal_table}, the system response or transfer function was quantified.
The weighted mean of the bracketing transfer functions was then used to scale 
the observed target squared visibility appropriately to produce measurements 
of $V^2$. This procedure resulted in a total of 35 estimations of $V^2$ and 
a summary of each night is given in Table~\ref{obs_table}.

It should be noted that the uncertainties in the calibrator angular diameters,
when compared to the measurement uncertainty, have a negligible effect on 
the final $V^2$ but for consistency, were included in the uncertainty 
calculation.

\begin{table}
\centering
    \caption{Adopted parameters of calibrators used during observations.}
    \label{cal_table}
    \begin{tabular}{@{}cccccc@{}}
	\hline
    HR 	 & Name 	& Spectral & V    & UD Diameter       & separation\\
	 &		&   Type   &      &  (mas)    	      & from $\beta$ Hyi\\
	\hline
    0705 & $\delta$ Hyi   & A3V    & 4.08 & $0.50 \pm 0.03$ &  11\fdg85 \\
    0806 & $\epsilon$ Hyi & B9V    & 4.10 & $0.39 \pm 0.03$ &  13\fdg05 \\
    7590 & $\epsilon$ Pav & A0V    & 3.96 & $0.46 \pm 0.05$ &  16\fdg53 \\
	\hline
    \end{tabular}
\end{table}

\begin{table*}
  \centering
  \begin{minipage}{120mm}
    \caption{Summary of observational data.  The night of the observation is 
             given in Columns 1 and 2 as a calendar date and a mean MJD. 
	     The nominal and mean projected baseline in units of metres is
	     given in Columns 3 and 4 respectively. The mean squared visibility,
	     standard deviation and number of observations during a night is 
	     given in the last three columns.}
    \label{obs_table}
    \begin{tabular}{@{}llcclccc}
	\hline
Date  	& MJD	& Nominal & Mean Projected  & Calibrators & ${\bar V^2}$ & ${\bar \sigma}$ & \# V$^2$\\	
	& 	& Baseline & Baseline  &	     &	& &\\	
	\hline
2004 October 08   & 53286.56 & 60 & 40.87 & $\epsilon$ Pav, $\delta$ Hyi, $\epsilon$ Hyi & 0.359 & 0.019 & 2 \\
2004 October 29   & 53307.48 & 5  &  3.40 & $\epsilon$ Pav, $\delta$ Hyi 	           & 1.037 & 0.033 & 4 \\
2005 September 20 & 53633.64 & 40 & 27.08 & $\epsilon$ Pav, $\delta$ Hyi                 & 0.671 & 0.021 & 6 \\
2005 November 11  & 53685.45 & 80 & 54.40 & $\epsilon$ Pav, $\delta$ Hyi, $\epsilon$ Hyi & 0.145 & 0.005 & 6 \\
2005 November 12  & 53686.42 & 80 & 54.23 & $\epsilon$ Pav, $\delta$ Hyi, $\epsilon$ Hyi & 0.141 & 0.009 & 5 \\
2005 November 13  & 53687.47 & 80 & 54.29 & $\epsilon$ Pav, $\delta$ Hyi, $\epsilon$ Hyi & 0.146 & 0.006 & 7 \\
2005 November 27  & 53701.43 & 5  & 3.40  & $\epsilon$ Pav, $\delta$ Hyi, $\epsilon$ Hyi & 0.983 & 0.024 & 5 \\
     \hline
    \end{tabular}
 \end{minipage}    
\end{table*}

%
%

\section{Angular Diameter}
\label{fit}

The brightness distribution of a star can be modeled, in the simplest case, 
to be a disc of uniform irradiance with angular diameter $\theta_{\rm UD}$. The 
theoretical response of a two aperture interferometer to such a model is 
given by
\begin{equation}
\label{udisc_v2}
|V|^2 = \left |\frac{2 J_1(\pi |\bmath{b}| \theta_{\rm UD} / \lambda)}
            {\pi |\bmath{b}| \theta_{\rm UD} / \lambda} \right|^2,
\end{equation}
where $J_1$ is a first order Bessel function, $\bmath{b}$ is the baseline
projected onto the plane of the sky and $\lambda$ is the observing wavelength.
For stars that have a compact atmosphere only small corrections are needed
to account for monochromatic limb-darkening -- these corrections can be 
found in \citet*{DTB2000}. 

An additional parameter, $A$, is included in the uniform disc model to 
account for instrumental effects arising from the differing spectral types
of \bhyi{} and the calibrators. Equation (\ref{udisc_v2}) 
becomes
\begin{equation}
\label{udisc2_v2}
|V|^2 = \left |A\frac{2 J_1(\pi |\bmath{b}| \theta / \lambda)}
            {\pi |\bmath{b}| \theta / \lambda} \right|^2.
\end{equation}

The estimation of $\theta_{\rm UD}$ and $A$ was completed using 
$\chi^2$ minimisation with an implementation of the Levenberg-Marquardt 
method to fit equation (\ref{udisc2_v2}) to all measures of $V^2$.
The formal uncertainties derived from the diagonal elements of the covariance 
matrix (which are calculated as part of the $\chi^2$ minimisation) were
verified by {\em Monte Carlo} simulations --
the visibility measurement errors may not strictly conform to a normal 
distribution and equation (\ref{udisc2_v2}) is non-linear. These
simulations subjected the model visibilities of each observation to a
Monte Carlo realisation of the (assumed) normal measurement error 
distribution to produce synthetic data sets. Estimates of the model 
parameters are then found using the synthetic datasets, thus building a
distribution of each parameter. 

\begin{figure}
\includegraphics[width=\linewidth]{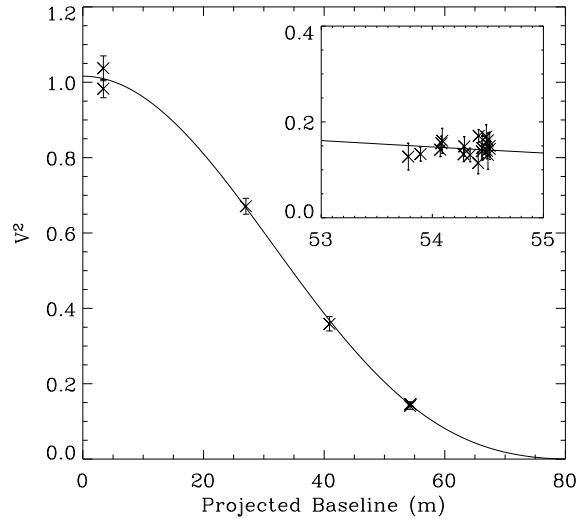}
\caption{\label{fig.vis} Nightly mean $V^2$ measures with
the fitted uniform disc overlayed. Inset: all $V^2$ measures at the 
(nominal) 80\,m SUSI baseline.}
\end{figure}

The reduced $\chi^2$ of the fit was 1.29 implying that the measurement 
uncertainties were underestimated. We have therefore scaled the
uncertainties in the fit parameters by $\sqrt{1.29}$ to obtain a
reduced $\chi^2$ of unity. Final values of $\theta_{\rm UD}$ and $A$ 
(with the associated $1\sigma$ parameter uncertainty) are 
$2.156\pm0.017$\,mas and $1.008\pm0.007$  respectively. The data, with 
the fitted uniform disc model overlaid, are
shown in Figure~\ref{fig.vis}. The fitted value of $A$ is  close
to unity indicating that the effect of the differing spectral types of 
the stars is very small and can be neglected. Furthermore, the 
resolution of SUSI during observations was such that the calibrator stars 
were essentially unresolved.

Analysis of the effect of the wide observing band was completed using
a G2~IV flux distribution (Davis et al.\ in preparation). 
Wide bandwidth effects 
are only significant when the interferometer's coherent field-of-view is 
smaller than the angular extent of source \citep{T+D2002}.
The coherent field-of-view during observations was found to be greater than 
7\,mas hence {\em bandwidth smearing} can be considered negligible. However, 
the interferometer's {\em effective} wavelength when observing a G2~IV star 
is approximately $696.6\pm 2.0$\,nm (Davis et al.\ in preparation). 
Subsequently, the final fit to 
equation (\ref{udisc2_v2}) was completed with the new effective wavelength.
Using the work of \citet{DTB2000}, the limb-darkening correction was 
determined to be 1.047 using the following parameters:  
$T_{\rm eff} = 5872$\,K, $\log g = 3.95$ and [Fe/H] = -0.17. These
parameters (with the exception of [Fe/H]) are the result of an iterative 
procedure whereby initial values, obtained from SIMBAD, were refined by 
subsequent fundamental parameter determination (see Section~\ref{sec:disc}). 
We note, however, that the change in $T_{\rm eff}$ and $\log g$ only resulted 
in a decrement of the last digit in the limb-darkening correction.
The parameter [Fe/H] is the mean of the values listed on SIMBAD and
we estimate an uncertainty in the limb-darkening correction of 0.002.
Therefore the limb-darkened diameter is $\theta_{\rm LD} = 2.257\pm0.019$\,mas  
and carries the caveat that the Kurucz models that \citet{DTB2000} based 
their work upon are accurate for this star.

%
%

\section{Discussion}
\label{sec:disc}

\begin{figure}
\centering
\includegraphics{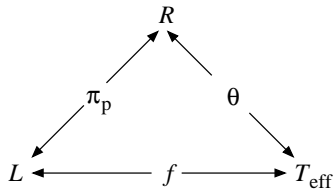}
\caption{\label{fig.params} Relationships between the observable quantities
(inner) and fundamental stellar parameters at the vertices.}
\end{figure}

Formulae to calculate the stellar luminosity ($L$), effective temperature 
($T_{\rm eff}$) and radius ($R$) from the observable quantities 
angular diameter ($\theta$), bolometric flux ($f$) and parallax ($\pi_{\rm p}$) 
are:
\begin{equation}
L = 4 \pi f\frac{C^2}{\pi_{\rm p}^2},
\end{equation}
\begin{equation}
T_{\rm eff} = \left[ \frac{4f}{\sigma\theta^2}\right]^{1/4},
\end{equation}
\begin{equation}
R = 2\theta \frac{C}{\pi_{\rm p}},
\end{equation}
where $\sigma$ is the Stefan-Boltzmann constant and $C$ is the conversion 
from parsecs to metres. The relationship between these quantities can be 
visualised as Fig.~\ref{fig.params}, where the inner quantities are the 
observables and the vertices are the fundamental stellar parameters.

\begin{figure}
\includegraphics[width=\linewidth]{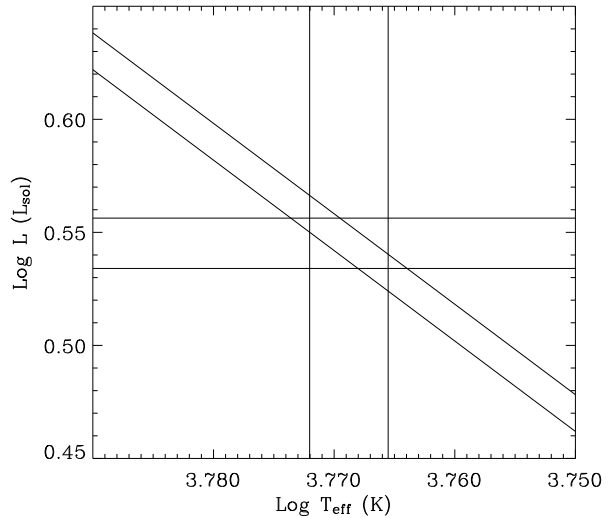}
\caption{\label{fig.hr} 
The 1-sigma constraints presented in this work on the effective temperature 
(vertical), luminosity (horizontal) and radius (diagonal), shown in an
H-R diagram.}
\end{figure}

Our measurement of the angular diameter of \bhyi, after correcting for limb
darkening, has an accuracy of 0.8\%.  Combining this with the {\em
Hipparcos\/} parallax (0.4\%) allows us to determine the radius of the star 
with an accuracy of 0.9\%, as given in Table~\ref{phys_table}.  This radius 
determination locates the star in the H-R diagram with much greater accuracy 
than was previously possible (see Fig.~\ref{fig.hr}) and will be extremely 
important for calculating theoretical models for \bhyi.  

Our angular diameter measurement, in combination with the bolometric flux 
of \citet{B+LG98} (the associated uncertainty estimation has been taken 
from \citealt{DiB98}) yields an effective temperature of $5872\pm44$\,K.
This value, with an accuracy of 0.7\%, is higher than those
presented in \citet{B+LG98} and \citet{DiB98}: $5710\pm29$\,K and
$5774\pm52$\,K respectively. While all three temperature estimations use the 
bolometric flux of \citet{B+LG98}, only our value uses a direct measurement 
of the angular diameter (and hence radius) of \bhyi{}. Furthermore, a
recent calibration of the MK system for late A-, F- and early G-type stars 
by \citet{GGH01} yields a value of $5850$\,K for a G2~IV star and the recent 
spectroscopic analysis by \citet{daSGP2006} produced an effective temperature
of $5964\pm70$\,K for \bhyi{}, both of which are consistent with our value.

The luminosity of \bhyi, found from the {\em Hipparcos\/} parallax
and bolometric flux, is $3.51\pm0.09$\,L$_{\odot}$ (following 
\citealt*{BPB2001}, we adopt L$_{\odot} = 3.842\times 10^{26}$\,W with an 
uncertainty of 0.4\%). The constraints these fundamental stellar values place
on the location of \bhyi{} in the H-R diagram are shown in Fig~\ref{fig.hr}. 

Calculating theoretical models for \bhyi{} is beyond the scope of this
paper. However, we are able to combine our radius measurement with the mean
density, inferred from the observed value of the large frequency separation
in the asteroseismic data, to determine the mass using:
\begin{equation} 
M = \frac{4}{3} \pi \bar{\rho} R^3.
\end{equation}
Taking the value for the mean density of \bhyi{}, determined to a precision 
of 0.6\% by \citet{BKA2007} using asteroseismology (see 
Table~\ref{phys_table}), we calculate the mass to be $1.07 \pm 0.03$\,M$_{\odot}$.
This is consistent with, but much more precise than, values in the 
literature that were estimated from modelling the position of \bhyi{} in the 
H-R diagram: $1.1$\,M$_{\odot}$ \citep{DLvdB98}, 
$1.10^{+0.04}_{-0.07}$\,M$_{\odot}$ \citep{F+M2003}, 
$1.17\pm0.05$\,M$_{\odot}$ \citep{daSGP2006}.
Finally, we can combine our radius with the mean density to produce an 
estimate of the surface gravity, 
\begin{equation}
g = \frac{4}{3}G\pi\bar{\rho}R,
\end{equation}
where G is the universal constant of gravitation. The value we obtain for
the surface gravity leads to $\log g = 3.952\pm0.005$, which has a
precision of 0.1\%. 

\begin{table}
    \caption{Physical parameters of \bhyi{}.}
    \label{phys_table}
    \begin{tabular}{@{}lccl@{}}
	\hline
	Parameter & Value  & Uncertainty (\%) & Source\\
	\hline
	$\theta_{\rm LD}$ (mas)   &$2.257\pm0.019$	   & 0.8 & this work\\
	$\pi_{\rm p}$ (mas) 	  &$133.78\pm0.51\ \ \ \ $ & 0.4 & ESA97   \\
	$f$ ($10^{-9}$\,Wm$^{-2}$)&$2.019\pm0.050$	   & 2.5 & BLG98,   \\		
				  &         		   &     & DiB98   \\
	\hline
	$L$ (L$_{\odot}$) &    $3.51\pm0.09$	    & 2.6 & this work \\
	$T_{\rm eff}$ (K) &    $5872\pm44\ \ \ \;$  & 0.7 & this work \\ 	
	$R$ (R$_{\odot}$) &   $1.814\pm0.017$       & 0.9 & this work \\
	\hline
	$\bar{\rho}$ ($\bar{\rho}_\odot$) & $0.1803\pm0.0011$ & 0.6 & BKA07 \\
	$M$ (M$_{\odot}$) 		  &   $1.07\pm0.03 $  & 2.8 & this work\\
	$\log g$          		  &  $3.952\pm0.005$  & 0.1 & this work\\
	\hline
    \end{tabular}
	\newline
	ESA97: \citet{ESA97}; BLG98: \citet{B+LG98};
	\newline 
	DiB98: \citet{DiB98}; BKA07: \citet{BKA2007} 	
\end{table}

%
%
\section{Conclusion}

We have presented the first angular diameter measurement of the G2 subgiant
\bhyi{}. In combination with literature values for the bolometric flux and
parallax, the angular diameter has constrained the stellar radius and 
effective temperature. The radius measurement, combined with the mean 
density determined from asteroseismology, allows the most accurate mass 
estimate of \bhyi{} to date. Indeed, this is perhaps the most precise mass 
determination of a solar-type star that is not in a binary system (apart 
from the Sun).

The constraints on $L$, $T_{\rm eff}$, $R$, $M$ and $\log g$ that we present 
will be invaluable in the future to critically test theoretical models
of \bhyi{} and its oscillations. As stressed by \citet{B+G94}, for example,
oscillation frequencies are of most importance for testing evolution theories
when the other fundamental stellar properties are well-constrained.

\section*{Acknowledgments}

This research has been jointly funded by The University of Sydney and the 
Australian Research Council as part of the Sydney University Stellar
Interferometer (SUSI) project. 
JRN and APJ acknowledge the support provided by a University of Sydney 
Postgraduate Award. APJ and SMO were supported by a Denison Postgraduate 
Award.
This research has made use of the SIMBAD database,
operated at CDS, Strasbourg, France.

\end{document}